\documentclass[reprint,aps,prl,twocolumn,superscriptaddress,floatfix,longbibliography]{revtex4-1}

\usepackage[dvips]{graphicx}
\usepackage{latexsym}
\usepackage{amsmath}
\usepackage{amsthm}
\usepackage{amsfonts}
\usepackage{amssymb}
\usepackage{mathptmx}
\usepackage{subfigure}
\usepackage{mathrsfs}

\usepackage{color}
\definecolor{darkblue}{rgb}{0,0.02,0.45}
\usepackage[colorlinks=true,citecolor=darkblue]{hyperref}

\usepackage[all]{hypcap}
\usepackage{gensymb}

\definecolor{cream}{RGB}{222,217,201}

\begin{document}

This manuscript has been authored by UT-Battelle,
LLC under Contract No. DE-AC05-00OR22725 with
the U.S. Department of Energy. The United States
Government retains and the publisher, by accepting
the article for publication, acknowledges that the
United States Government retains a non-exclusive, paidup,
irrevocable, world-wide license to publish or reproduce
the published form of this manuscript, or
allow others to do so, for United States Government
purposes. The Department of Energy will
provide public access to these results of federally
sponsored research in accordance with the DOE
Public Access Plan(http://energy.gov/downloads/doepublic-access-plan).
\clearpage

\title{Crystal and magnetic structures of magnetic topological insulators MnBi$_2$Te$_4$ and MnBi$_4$Te$_7$}

\author{Lei Ding}
\email{dingl@ornl.gov}
\affiliation{Neutron Scattering Division, Oak Ridge National Laboratory, Oak Ridge, TN 37831, USA}
\author{Chaowei Hu}
\affiliation{Department of Physics and Astronomy and California NanoSystems Institute, University of California, Los Angeles, CA 90095, USA}
\author{Feng Ye}
\author{Erxi Feng}
\affiliation{Neutron Scattering Division, Oak Ridge National Laboratory, Oak Ridge, TN 37831, USA}
\author{Ni Ni}
\affiliation{Department of Physics and Astronomy and California NanoSystems Institute, University of California, Los Angeles, CA 90095, USA}
\author{Huibo Cao}
\email{caoh@ornl.gov}
\affiliation{Neutron Scattering Division, Oak Ridge National Laboratory, Oak Ridge, TN 37831, USA}
\date{September 06, 2019}

\begin{abstract}
Using single crystal neutron diffraction, we present a systematic investigation of the crystal structure and magnetism of van der Waals topological insulators MnBi$_2$Te$_4$ and MnBi$_4$Te$_7$, where rich topological quantum states have been recently predicted and observed. Structural refinements reveal that considerable Bi atoms occupied on the Mn sites in both materials, distinct from the previously reported antisite disorder. We show unambiguously that MnBi$_{2}$Te$_{4}$ orders antiferromagnetically below 24 K featured by a magnetic symmetry $R_I$-${3c}$ while MnBi$_{4}$Te$_{7}$ is antiferromagnetic below 13 K with a magnetic space group $P_c$-${3c1}$. They both present antiferromagnetically coupled ferromagnetic layers with spins along the $c$-axis. We put forward a stacking rule for the crystal structure of an infinitely adaptive series MnBi$_{2n}$Te$_{3n+1}$ (n$\geq$1) with a building unit of [Bi$_2$Te$_3$]. By comparing the magnetic properties between MnBi$_{2}$Te$_{4}$ and MnBi$_{4}$Te$_{7}$, together with the recent density-functional theory calculations, we concluded that a two-dimensional magnetism limit might be realized in the derivatives. Our work may promote the theoretical studies of topological magnetic states in the series of MnBi$_{2n}$Te$_{3n+1}$.

\end{abstract}

\maketitle

\textit{Introduction.} In condensed matter physics, van der Waals (vdW) magnetic heterostructures stacked layer-by-layer in a controlled sequence have attracted a great deal of interest as they have been found to show exotic physical properties and emergent phenomena~\citep{Geim2007, ChengGong2017, BevinHuang2017, Tokura2019}. Novel properties in these materials can be controlled by tuning the stacked atomic layers, paving the way for designing new quantum materials \citep{Geim2007}. VdW magnetic topological insulators have been suggested as a promising material platform for the exploration of exotic topological quantum phenomena such as the quantum anomalous Hall effect (QAHE), Majorana fermions as well as topological magnetoelectric effect\citep{Mong2010, XiaogangWen2019, Tokura2019, JiahengLi2019}. However, a homogenous heterostructure with intrinsic magnetism, an ideal platform for studying such topological quantum effects, is experimentally elusive \citep{Tokura2019}.

Very recently, MnBi$_2$Te$_4$ was proposed to be the first intrinsic antiferromagnetic (AFM) topological insulator.\citep{YujieHao2019, DongqinZhang2019, Hirahara2017, YanGong2019, Otrokov2019, Otrokov2018, Zeugner2019, JiahengLi2019, BoChen2019, YujunDeng2019, SenghuatLee2018} It has been shown that below 24 K, MnBi$_2$Te$_4$ orders into an A-type magnetic structure based on magnetic properties, density-functional theory calculations and powder neutron diffraction measurement \citep{Zeugner2019, Otrokov2018, JiaqiangYan2019, JiahengLi2019}. Since such a spin configuration breaks the product ($S$) of the time-reversal symmetry and the primitive-lattice translational symmetry at the (001) surface, it is expected that a gapped surface Dirac cone can be observed by the angle-resolved photoemission spectroscopy (ARPES) made on the cleaved (001) surface\citep{Otrokov2019, Mong2010}. However, both gapped and gapless \citep{Otrokov2018, YanGong2019, YujieHao2019, BoChen2019} Dirac cones have been observed in ARPES measurements. This casts doubt on the magnetic configuration determined using the powder neutron diffraction data \citep{JiahengLi2019}. Furthermore, a new family of MnBi$_{2n}$Te$_{3n+1}$ ($n=$ 2, 3) were later discovered \cite{Aliev2019, Souchay2019}. Among them, MnBi$_4$Te$_7$ \cite{ChaoweiHu2019, Vidal2019, Aliev2019} and MnBi$_6$Te$_{10}$ \citep{Aliev2019} have been suggested to be new magnetic topological insulators with weak interlayer magnetic coupling through a combined ARPES and first-principles calculations \citep{ChaoweiHu2019}. Although the A-type AFM was suggested for MnBi$_4$Te$_7$ by its anisotropic magnetic properties, no neutron experiment has been reported to prove it. Another riddle in both MnBi$_2$Te$_4$ and MnBi$_4$Te$_7$ is the saturated magnetic moment that is found to be about 3 $\mu_B$ in both cases \citep{SenghuatLee2018, BoChen2019, Vidal2019, ChaoweiHu2019}. This is significantly smaller than 5 $\mu_B$ expected for the Mn$^{2+}$ ion. To elucidate this reduction and unambiguously determine their magnetic structures, single crystal neutron diffraction experiment, which can map out the complete magnetic reflections with a better resolution, is urged.

Defects are of great importance in optimizing and understanding the surface states and magnetism in magnetic topological insulators \citep{Cava2013,Hor2010}. Antisite defects in MnBi$_{2n}$Te$_{3n+1}$ have been found to be divergent. For example, by combining single-crystal x-ray diffraction and electron microscopy, Zeugner $et\ al.$ have shown the presence of antisite disorder between Mn and Bi sites and Mn vacancies in the nominal MnBi$_2$Te$_4$ which yield the composition Mn$_{0.85(3)}$Bi$_{2.10(3)}$Te$_4$ \citep{Zeugner2019} while Yan $et\ al.$ later found the cationic disorder with only about 3\% of the Bi sites occupied by Mn estimated by scanning tunnelling microscopy \citep{JiaqiangYan2019}. Similar defects have been reported in the nominal MnBi$_4$Te$_7$ by x-ray diffraction and transmission electron microscopy \citep{Souchay2019, Vidal2019}. However, in light of the great difference of Mn and Bi in electronegativity, antisite defects are basically unfavourable. To rationally sort out the debate, it is natural to employ neutron diffraction technique as Mn and Bi atoms have opposite sign of the neutron scattering length. 

Beyond the investigation of the magnetic structures, motivated by the above mentioned discoveries, our work is devoted to setting up the relationship between crystal structure and magnetic properties by examining the simple lattice-stacking rule starting from the prototype topological insulator Bi$_2$Te$_3$ \cite{HaijunZhang2009,YulinChen2009}. Conventionally, its structure is described by the stacking of three ``quintuple layer'' building blocks \cite{Atuchin2012, Cava2013}. The bond coupling is rather strong between two atomic layers within one quintuple layer but much weaker, predominantly of the vdW type, between neighboring quintuple layers. Yet, ``quintuple layer'' stacking description does not reflect the structural symmetry of Bi$_2$Te$_3$ that may be important in understanding the topological properties in MnBi$_{2n}$Te$_{3n+1}$. The proposed stacking rule for building magnetic topological insulator Mn-Bi-Te series in this work practically captures the structural symmetry, reflects the precise layer stacking sequence, and reveals an infinitely adaptive series, stimulating theoretical calculations of exotic quantum states in the series of MnBi$_{2n}$Te$_{3n+1}$.

\begin{figure}
\centering
\includegraphics[width=1\linewidth]{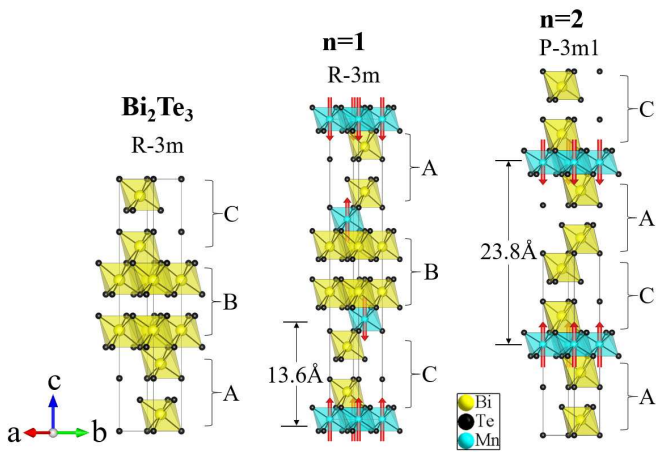}
\caption{Schematic drawings of the crystal structures of Bi$_2$Te$_3$ and MnBi$_{2n}$Te$_{3n+1}$ ($n=$ 1, 2). Note that for the latter cases, spin arrangements are denoted by red arrows.}\label{fig:1}
\end{figure}

\begin{figure*}
\centering
\includegraphics[width=0.9\linewidth]{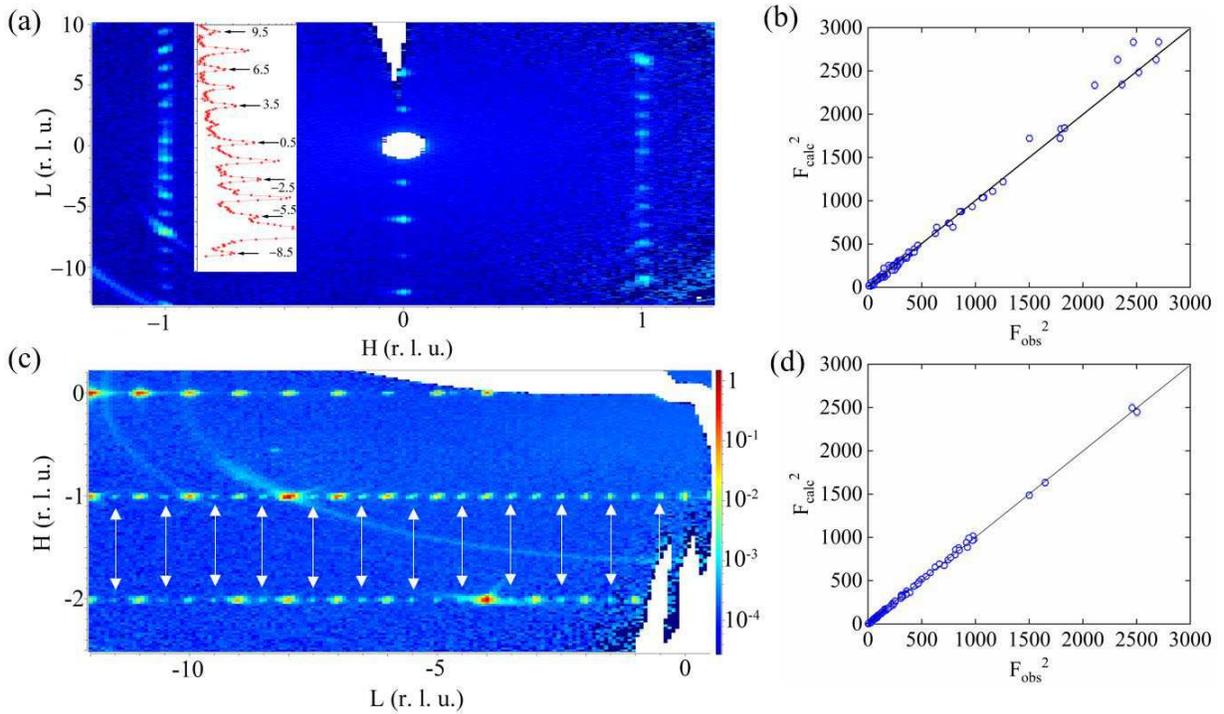}
\caption{(a) Contour map of neutron intensity of MnBi$_2$Te$_4$ in the (H 0 L) reciprocal plane at 7\,K measured on the CORELLI. (b) The results for the nuclear and magnetic structure refinements of MnBi$_2$Te$_4$ at 4.5 K with neutron data from the DEMAND. (c) Contour map of neutron intensity of MnBi$_4$Te$_7$ in the (H 0 L) reciprocal plane at 7\,K measured on the CORELLI. Arrows mark the magnetic reflections. (d) The results for the nuclear and magnetic structure refinements of MnBi$_4$Te$_7$ at 4.5 K with neutron data from the DEMAND}\label{fig:3}
\end{figure*}

\textit{Crystal structures of MnBi$_2$Te$_4$ and MnBi$_4$Te$_7$.} We examined the crystal structure of both powder and single crystalline MnBi$_2$Te$_4$ and MnBi$_4$Te$_7$ samples using x-ray and neutron diffraction techniques \citep{FengYe2018}. For MnBi$_2$Te$_4$, the slice view of neutron diffraction data at 7 K is shown in Fig.\ref{fig:3}(a) while the powder x-ray diffraction pattern is presented in \citep{suppl}. For MnBi$_4$Te$_7$, Fig. \ref{fig:3}(c) shows the contour map of neutron diffraction at 7 K, where the presence of sharp reflections indicates the high quality of the crystal. Single crystal neutron diffraction was also performed on the HB-3A DEMAND \citep{HB3A, DEMAND} at room temperature on the same crystals \citep{suppl}. We have refined both the single crystal neutron diffraction data from the DEMAND and x-ray powder diffraction data using the Fullprof software \citep{Fullprof1993}. The refinement results as well as structural parameters are shown in Fig. \ref{fig:3} and Table S1-S4 \cite{suppl}, respectively. The refined lattice parameters are presented in Table~\ref{tab:1}, which are in good agreement with the previous reports~\citep{DongsunLee2013, Aliev2019}. Our refinement shows that about 18(1) \% of Bi occupied at the Mn sites in the as-grown crystal MnBi$_2$Te$_4$ whereas there is negligible Mn 1(1) \% resided on the Bi sites \citep{suppl}. This sort of non-stoichiometry effect becomes more robust in the MnBi$_4$Te$_7$ crystal where there is 27.9(4) \% of Bi on the Mn sites. Such results are substantially different from the intermixing between Mn and Bi sites and vacancies of Mn in MnBi$_2$Te$_4$ and MnBi$_4$Te$_7$ documented previously \citep{Zeugner2019, Otrokov2018, Souchay2019, Vidal2019, JiaqiangYan2019}. 

\begin{figure}
\centering
\includegraphics[width=1\linewidth]{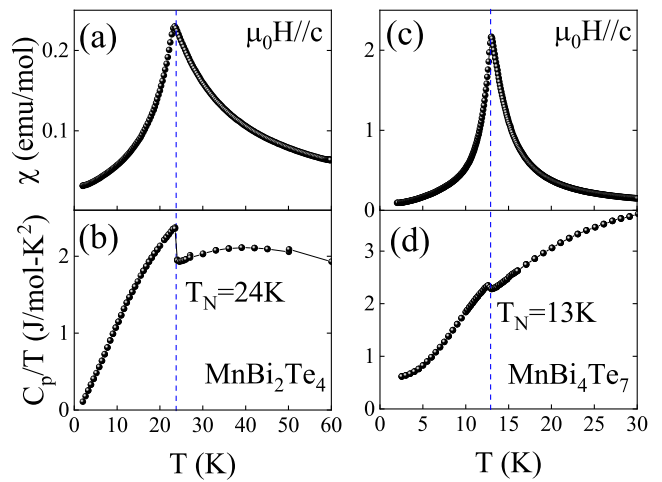}
\caption{Temperature dependence of the magnetic susceptibilities of MnBi$_2$Te$_4$ (a) and  MnBi$_4$Te$_7$ (c) down to 2\,K with magnetic field of 100 Oe under zero-field-cooling condition. Zero-field specific heat of  MnBi$_2$Te$_4$ (b) and  MnBi$_4$Te$_7$ (d) down to 2\,K.}  \label{fig:2}
\end{figure}

\textit{Thermodynamic properties of MnBi$_{2}$Te$_{4}$ and MnBi$_{4}$Te$_{7}$.} The ZFC temperature dependent magnetic susceptibilities of MnBi$_2$Te$_4$ and MnBi$_4$Te$_7$ measured with the field parallel to the $c$-axis are shown in Fig.~\ref{fig:2}. For the MnBi$_2$Te$_4$ case, the susceptibility first increases with decreasing temperature and exhibits a sharp cusp at T$_N$ = 24 K, signaling the AFM ordering. In order to further characterize the magnetic phase transition, we measured the specific heat of MnBi$_2$Te$_4$. As shown in Fig. \ref{fig:2}(b), the cusp at 24\,K is indicative of the AFM transition, in good agreement with the magnetic susceptibility data. Thus MnBi$_2$Te$_4$ only undergoes one AFM transition at 24 K.
The magnetic susceptibility of MnBi$_4$Te$_7$ shows a sharp peak at T$_N$=13\,K (Fig.\ref{fig:2}(c)), indicating an AFM transition. The nature of the transition can be crosschecked by the specific heat data of MnBi$_4$Te$_7$. As shown in Fig.\ref{fig:2}(d), the specific heat shows a small anomaly at 13\,K, consistent with the magnetic susceptibility results and the previous results.\citep{ChaoweiHu2019, Vidal2019} The Ne\'el temperature of MnBi$_4$Te$_7$ is about two times smaller than that of MnBi$_2$Te$_4$.

\textit{A-type magnetic structure of MnBi$_{2}$Te$_{4}$ and MnBi$_{4}$Te$_{7}$.} The magnetic structure of MnBi$_2$Te$_4$ has been previously investigated using powder neutron diffraction~\cite{JiaqiangYan2019}. They found that the AFM structure has the magnetic symmetry ${P_c}$-${3c1}$ (BNS symbol) propagated by a vector \textbf{k}=(0, 0, 1/2), which is a lower magnetic symmetry than that determined from our neutron data. As shown in Fig.~\ref{fig:3}(a), our neutron diffraction experiment reveals that a set of magnetic reflections, which appears at 7 K, can be indexed by a vector \textbf{k}=(0, 0, 3/2). As a matter of fact, the magnetic reflections shown in Ref.~\citep{JiaqiangYan2019} should be reasonably indexed by this vector, implying that the proposed magnetic space group should be reconsidered here. To solve the magnetic structure of MnBi$_2$Te$_4$, we have carried out the magnetic symmetry analysis by considering the \textbf{k} vector and the parent space group ${R}$-${3m1'}$ with help of the ISODISTORT software \citep{ISODISTORT} and Bilbao Crystallography Server \citep{bilbao}. There are two active magnetic irreducible representations, mT2+ and mT3+. After testing these candidates, we found the irrep mT2+, corresponding to the magnetic space group $R_I$-$3c$, is the only solution while the latter irrep mT3+ conveying three different magnetic structures is incompatible with our neutron data \citep{suppl}. Magnetic neutron diffraction data measured at 4.5 K on the DEMAND were then refined using the magnetic symmetry $R_I$-$3c$ and the results of the refinement is shown in Fig. \ref{fig:3}(b). It turns out that spins line up ferromagnetically in the ${ab}$ plane below 24 K whereas between layers, magnetic moments are antiparallel. The refined magnetic moment of Mn$^{2+}$ at 4.5 K is 4.7(1)$\mu_B$, in good accordance with the expected totally ordered moment 5$\mu_B$ for Mn$^{2+}$ ion.

Single crystal neutron diffraction data of MnBi$_4$Te$_7$ show clearly the appearance of magnetic reflections at the (H 0 L+0.5) positions (H, L denote the Miller index) upon cooling below 13 K, confirming the formation of a long-range AFM order. As shown in Fig.~\ref{fig:3}(c) all magnetic reflections can be indexed by the propagation vector \textbf{k}=(0, 0, 1/2). Starting with the parent space-group $P$-${3m1'}$ and the propagation vector \textbf{k} in A point in the Brillouin zone, through ISODISTORT, two active magnetic irreducible representations, mA1- and mA3- were obtained. We found that magnetic space group ${P_c}$-${3c1}$ (BNS symbol, basis={(1,0,0),(0,1,0),(0,0,2)}, origin=(0, 0, 1/2)), generated from the single active mA1-, can be adopted to describe the magnetic structure. Magnetic structure model was refined on the neutron data collected at 4.5 K on the DEMAND. The refined magnetic structure is characteristic of an AFM arrangement between the adjacent FM layers, as shown in Fig~\ref{fig:1}. The refined total magnetic moment at 4.5 K is 4.01(9)$\mu_B$ along the $c$-axis, relatively smaller than the value of 5 $\mu_B$ expected for S = 5/2 of Mn$^{2+}$. Here, the small discrepancy is likely because it is not fully ordered yet at 4.5 K, indicated by the ordering parameter shown below. It appears that MnBi$_4$Te$_7$ bears a similar spin arrangement between layers (different magnetic symmetry) to MnBi$_2$Te$_4$ but magnetically orders at a much lower temperature due to the increased distance between magnetic Mn-layers.

The seemingly reduced saturated magnetic moment in both MnBi$_{2}$Te$_{4}$ and MnBi$_{4}$Te$_{7}$ can be reasonably understood by our neutron diffraction results. The deficiency of Mn on the Mn sites in both cases naturally explains the small saturated magnetic moments detected by the bulk magnetization. As determined by the neutron diffraction, the ordered magnetic moment of Mn is reasonably close to the theoretical value.     

\begin{figure}
\centering
\includegraphics[width=1\linewidth]{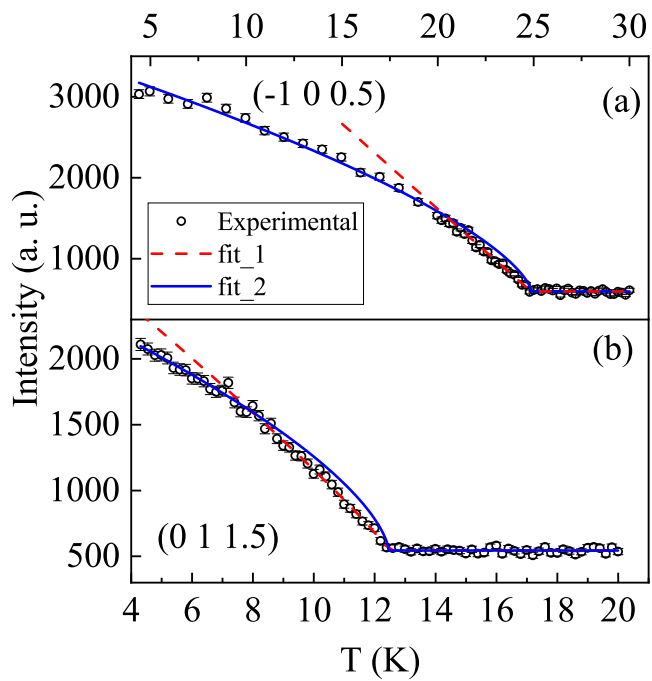}
\caption{Magnetic order parameter upon warming at the magnetic reflections (-1 0 0.5) and (0 1 1.5) for MnBi$_2$Te$_4$ (a) and MnBi$_4$Te$_7$ (b) measured at DEMAND, respectively. Dashed and solid lines represent the results of a power-law fits in different temperature regions.}\label{fig:4}
\end{figure}
   
\begin{table*}[t]
\caption{Derivatives of the vdW MnBi$_{2n}$Te$_{3n+1}$ topological insulators. Lattice parameters (i: experiment, ii: prediction) are shown at room temperature. The lattice parameters for each sequence of the stacking units are based on Bi$_2$Te$_3$: a=b=4.3896(2)\,\r A, c=30.5019(10)\,\r A \label{tab:1}\citep{Atuchin2012}. When $n$=2+3$m$ ($m=$ 0, 1, 2...), the lattice of derivatives is primitive. The predicted magnetic space groups (MSG) are made based on a symmetry analysis using the \textbf{k} vectors of (0, 0, 3/2) for the rhombohedral lattice, (0, 0, 1/2) for the hexagonal lattice, (0, 0, 0) for the cases with ferromagnetic (FM) order.}
\begin{tabular}{ccccccccccc}
\hline
\hline
\multicolumn{1}{c}{n}   & &  \multicolumn{1}{c}{Sequence} & &  \multicolumn{1}{c}{SG} &  & \multicolumn{2}{c}{Lattice parameter} & & \multicolumn{1}{c}{ Magnetic properties and MSG} \\
 \hline
 1  & &  \textit{M}-C-\textit{M}-B-\textit{M}-A-\textit{M} & & ${R}$-${3m}$ & & 4.3336(2) \AA & 40.926(3) \AA [i, this work]   & & AFM, $R_I$-${3c}$ [i, this work]\\
 2  & &  A-\textit{M}-C & & ${P}$-${3m1}$ & & 4.3453(5) \AA & 23.705(3) \AA [i, this work]   & & AFM, ${Pc}$-${3c1}$ [i, this work]\\
   & &   & &  & &  & 23.81 \AA [ii]   & & \\
 3  & &  \textit{M}-C-A-B-\textit{M}-A-B-C-\textit{M}-B-C-A-\textit{M} & & ${R}$-${3m}$ & & 4.3745(3) \AA & 101.985(8) \AA [i]\cite{Aliev2019}   & & AFM, $R_I$-${3c}$ / FM, ${R}$-${3m}^{\prime}$[ii]\\
    & &   & &  & &  & 101.94 \AA [ii]   & & \\
 4  & &  \textit{M}-C-A-B-C-\textit{M}-B-C-A-B-\textit{M}-A-B-C-A-\textit{M} & & ${R}$-${3m}$ & &  & 132.45 \AA [ii]  & & AFM, $R_I$-${3c}$ / FM, ${R}$-${3m}^{\prime}$[ii] \\
 5  & &  \textit{M}-C-A-B-C-A-\textit{M} & & ${P}$-${3m1}$ & &  &  54.32 \AA [ii]  & & AFM, ${Pc}$-${3c1}$ / FM, ${P}$-${3m}^{\prime}1$[ii] \\
 6  & &  \textit{M}-C-A-B-C-A-B-\textit{M}-A-B-C-A-B-C-\textit{M}-B-C-A-B-C-A-\textit{M} & & ${R}$-${3m}$ & &  & 193.47 \AA [ii]  & & AFM, $R_I$-${3c}$ / FM, ${R}$-${3m}^{\prime}$[ii]\\
\hline
\hline
\end{tabular}
\end{table*}

Having known the magnetic structures of the vdW magnets with $n=1$, 2, we now examine the temperature dependence of the strongest magnetic reflections for each case. The temperature-dependent intensity of the magnetic reflections ( -1 0 0.5) and (0 1 1.5) for MnBi$_2$Te$_{4}$ and MnBi$_4$Te$_{7}$, respectively are shown in Fig.~\ref{fig:4}. They may follow an empirical power-law behavior \citep{Brush1967, Birgeneau1970},
\begin{equation}
I=A(\dfrac{T_M-T}{T_M})^{2\beta}+B
\end{equation}
where T$_M$ is the critical temperature for magnetic phase transitions, A is a proportionality constant, $\beta$ is the order parameter critical exponent and B is the background. Fits to the power law were performed for two temperature regions for the two cases. The best fit in the temperature ranges of 21-30 K and 9-20 K yields the Ne\'el temperatures T$_N$=24.8(1)\,K and 12.5(1)\,K, the critical exponents $\beta$=0.50(5) and 0.45(3) for MnBi$_2$Te$_{4}$ and MnBi$_4$Te$_{7}$, respectively. Both critical exponents determined in the two temperature regions that are within the critical regions are in accordance with that of the Ginzburg-Landau theory. In the temperature range of 4.5-21 K, the best fit for MnBi$_2$Te$_{4}$ yields the critical exponent $\beta$= 0.32(1), a value relatively smaller than the results in Ref.\citep{JiaqiangYan2019}. This $\beta$ value is in fact very close to the value, 0.325, expected for a universality class of the three-dimensional Ising model\citep{Brush1967}. In the lower temperature range 4.5 K $<$ T $<$ 9 K, the critical exponent $\beta$=0.32(2) of MnBi$_4$Te$_{7}$ is very close to that of MnBi$_2$Te$_{4}$. This seems to be contradictory to the less three-dimensional magnetism in MnBi$_4$Te$_{7}$ caused by the much larger interlayer Mn-Mn distance. The reason may lie in the fact that the fitting for the MnBi$_4$Te$_{7}$ case was done in a temperature range that is relatively close to the critical region which often comes with a crossover from three-dimensional to two-dimensional behavior upon cooling \citep{Birgeneau1970, Wildes2006}.

\textit{Stacking rule and crystal structures of vdW MnBi$_{2n}$Te$_{3n+1}$.}
Based on the crystal structure of MnBi$_2$Te$_4$, MnBi$_4$Te$_7$, as shown in Fig.\ref{fig:1}, we can put forward a stacking rule of the crystal structure of this family that is based on the rhombohedral lattice of Bi$_2$Te$_3$. The aforementioned structural description of the Bi$_2$Te$_3$ is based on the stacking of three ``quintuple layer" building blocks. This description has been used to predict the infinitely adaptive series of thermoelectric materials \cite{Bos2007, Cava2013}. However, it seemingly does not capture the essential structural symmetry but rather simply depicts the stacking lattice. As shown in Fig.~\ref{fig:1}, the crystal structure of Bi$_2$Te$_3$ is designated as A-B-C stacking sequence with A, B and C representing distinct [Bi$_2$Te$_3$] units. Imposed by the inversion and rhombohedral centering translation, the A unit can be progressed into B, and subsequently C configuration. Simply, we introduce magnetic atoms by inserting M (M denotes MTe$_6$ octahedra layer) layers into this A-B-C stacking sequence, giving rise to an infinitely adaptive series MnBi$_{2n}$Te$_{3n+1}$. Accordingly, we can immediately name the sequence $M$-C-$M$-B-$M$-A-$M$ with $n=1$ which preserves the crystal symmetry $R$-${3m}$. This yields the compound MnBi$_2$Te$_4$, as illustrated in Fig. \ref{fig:1}. With $n=2$, where one more [Bi$_2$Te$_3$] unit is added, we get the stacking sequence A-$M$-C with the space group $P$-$3m1$. It corresponds to the compound MnBi$_4$Te$_7$, indicating a great compatibility between these structural units. It occurs that the B-type unit occurring in MnBi$_2$Te$_4$ disappears in MnBi$_4$Te$_7$. Consequently, this leads to a general stacking rule for a vast number of vdW magnets in this family: the magnetic Mn-layer can replace one of A, B or C-type unit but still leaves the -A-B-C- stacking sequence unchanged. Indeed, with the increment of the value of $n$ ($n \geq$ 1), the sequence of the building units, crystal symmetry and lattice parameters of the corresponding vdW magnet can be generated and listed in Table~\ref{tab:1}. Following this stacking rule, one can make an infinitely adaptive series as those listed in Table~\ref{tab:1} (here we only list up to $n=6$).

\textit{Magnetism of MnBi$_{2n}$Te$_{3n+1}$ derivatives.} 
Recently, theoretical calculations on MnBi$_2$Te$_4$ \citep{Otrokov2019} have shown that in the $(ab)$ plane the FM interaction between the first nearest neighbors J$_1$=1.693 meV strongly dominates over all others. By considering all the exchange interactions in the layer and magnetic anisotropy energy, they suggest the FM ordering temperature of T$_c$=12(1)K in a single free-standing MnBi$_2$Te$_4$ septuple layer \citep{Otrokov2019}. Interestingly, MnBi$_4$Te$_7$ orders antiferromagnetically at 13 K, very close to the aforementioned 12(1) K. This indicates that in MnBi$_4$Te$_7$ intralayer exchange interaction is robust whereas the interlayer one is minimal. Indeed, recent calculations on MnBi$_4$Te$_7$ yielded that J$_1$ and J$_{\perp}$ (a summed exchange coupling between the adjacent layers) \citep{ChaoweiHu2019, Vidal2019} are 1.704 meV and -0.150 meV, respectively. With the increase of non-magnetic building units, FM could be realized by quenching all interlayer exchange interactions. Therefore, for MnBi$_{2n}$Te$_{3n+1}$ with n$>$2, if we assume their magnetic ground state to be either A-type AFM or FM with all magnetic moments being aligned along the $c$-axis, the magnetic space group for each compound is predicted and presented in Table \ref{tab:1}. Experiments to characterize other derivatives and to ascertain the predicted magnetic space groups are called.

\textit{Conclusion.} In summary, we have investigated systematically the crystal structure and magnetism of vdW topological insulators MnBi$_{2}$Te$_{4}$ and MnBi$_{4}$Te$_{7}$. Considerable Mn deficiency has been found on the Mn sites for both cases. Our results have shown that the ordered magnetic moments of MnBi$_{2}$Te$_{4}$ and MnBi$_{4}$Te$_{7}$ are reasonably close to the expected 5 $\mu_B$ for Mn$^{2+}$ ion. We have revealed a simple lattice-stacking rule that can lead to an infinitely adaptive series of MnBi$_{2n}$Te$_{3n+1}$. Although different magnetic symmetry exists in MnBi$_{2}$Te$_{4}$ and MnBi$_{4}$Te$_{7}$ with the former being $R_I$-${3c}$ and the latter being $P_c$-${3c1}$, our single crystal neutron diffraction measurements have unambiguously established that in the ordered state, both compounds show the A-type AFM structure with all spins parallel in the $ab$ plane but antiparallel along the $c$ axis which is critical for the observation of quantized anomalous Hall effect when they are exfoliated down to a few layers.

\textit{Acknowledgements.}
We thank C. D. Batista for useful discussions. The research at Oak Ridge National Laboratory (ORNL) was supported by the U.S. Department of Energy (DOE), Office of Science, Office of Basic Energy Sciences, Early Career Research Program Award KC0402010, under Contract DE-AC05-00OR22725 and the U.S. DOE, Office of Science User Facility operated by the ORNL. Work at UCLA was supported by the U.S. Department of Energy (DOE), Office of Science, Office of Basic Energy Sciences under Award Number DE-SC0011978. The US Government retains, and the publisher, by accepting the article for publication, acknowledges that the US Government retains a nonexclusive, paid-up, irrevocable, worldwide license to
publish or reproduce the published form of this manuscript,
or allow others to do so, for US Government purposes. The
Department of Energy will provide public access to these
results of federally sponsored research in accordance with the
DOE Public Access Plan \citep{DOE}.


\begin{thebibliography}{40}%
\makeatletter
\providecommand \@ifxundefined [1]{%
 \@ifx{#1\undefined}
}%
\providecommand \@ifnum [1]{%
 \ifnum #1\expandafter \@firstoftwo
 \else \expandafter \@secondoftwo
 \fi
}%
\providecommand \@ifx [1]{%
 \ifx #1\expandafter \@firstoftwo
 \else \expandafter \@secondoftwo
 \fi
}%
\providecommand \natexlab [1]{#1}%
\providecommand \enquote  [1]{``#1''}%
\providecommand \bibnamefont  [1]{#1}%
\providecommand \bibfnamefont [1]{#1}%
\providecommand \citenamefont [1]{#1}%
\providecommand \href@noop [0]{\@secondoftwo}%
\providecommand \href [0]{\begingroup \@sanitize@url \@href}%
\providecommand \@href[1]{\@@startlink{#1}\@@href}%
\providecommand \@@href[1]{\endgroup#1\@@endlink}%
\providecommand \@sanitize@url [0]{\catcode `\\12\catcode `\$12\catcode
  `\&12\catcode `\#12\catcode `\^12\catcode `\_12\catcode `\%12\relax}%
\providecommand \@@startlink[1]{}%
\providecommand \@@endlink[0]{}%
\providecommand \url  [0]{\begingroup\@sanitize@url \@url }%
\providecommand \@url [1]{\endgroup\@href {#1}{\urlprefix }}%
\providecommand \urlprefix  [0]{URL }%
\providecommand \Eprint [0]{\href }%
\providecommand \doibase [0]{http://dx.doi.org/}%
\providecommand \selectlanguage [0]{\@gobble}%
\providecommand \bibinfo  [0]{\@secondoftwo}%
\providecommand \bibfield  [0]{\@secondoftwo}%
\providecommand \translation [1]{[#1]}%
\providecommand \BibitemOpen [0]{}%
\providecommand \bibitemStop [0]{}%
\providecommand \bibitemNoStop [0]{.\EOS\space}%
\providecommand \EOS [0]{\spacefactor3000\relax}%
\providecommand \BibitemShut  [1]{\csname bibitem#1\endcsname}%
\let\auto@bib@innerbib\@empty
\bibitem [{\citenamefont {Geim}\ and\ \citenamefont
  {Novoselov}(2007)}]{Geim2007}%
  \BibitemOpen
  \bibfield  {author} {\bibinfo {author} {\bibfnamefont {A.~K.}\ \bibnamefont
  {Geim}}\ and\ \bibinfo {author} {\bibfnamefont {K.~S.}\ \bibnamefont
  {Novoselov}},\ }\bibfield  {title} {\enquote {\bibinfo {title} {The rise of
  graphene},}\ }\href {\doibase 10.1038/nmat1849} {\bibfield  {journal}
  {\bibinfo  {journal} {Nature Mater.}\ }\textbf {\bibinfo {volume} {6}},\
  \bibinfo {pages} {183} (\bibinfo {year} {2007})}\BibitemShut {NoStop}%
\bibitem [{\citenamefont {Gong}\ \emph {et~al.}(2017)\citenamefont {Gong} \emph
  {et~al.}}]{ChengGong2017}%
  \BibitemOpen
  \bibfield  {author} {\bibinfo {author} {\bibfnamefont {C.}~\bibnamefont
  {Gong}} \emph {et~al.},\ }\bibfield  {title} {\enquote {\bibinfo {title}
  {{Discovery of intrinsic ferromagnetism in two-dimensional van der Waals
  crystals}},}\ }\href {\doibase 10.1038/nature22060} {\bibfield  {journal}
  {\bibinfo  {journal} {Nature}\ }\textbf {\bibinfo {volume} {546}},\ \bibinfo
  {pages} {265--269} (\bibinfo {year} {2017})}\BibitemShut {NoStop}%
\bibitem [{\citenamefont {Huang}\ \emph {et~al.}(2017)\citenamefont {Huang}
  \emph {et~al.}}]{BevinHuang2017}%
  \BibitemOpen
  \bibfield  {author} {\bibinfo {author} {\bibfnamefont {B.}~\bibnamefont
  {Huang}} \emph {et~al.},\ }\bibfield  {title} {\enquote {\bibinfo {title}
  {Layer-dependent ferromagnetism in a van der waals crystal down to the
  monolayer limit},}\ }\href {\doibase 10.1038/nature22391} {\bibfield
  {journal} {\bibinfo  {journal} {Nature}\ }\textbf {\bibinfo {volume} {546}},\
  \bibinfo {pages} {270--273} (\bibinfo {year} {2017})}\BibitemShut {NoStop}%
\bibitem [{\citenamefont {Tokura}\ \emph {et~al.}(2019)\citenamefont {Tokura},
  \citenamefont {Yasuda},\ and\ \citenamefont {Tsukazaki}}]{Tokura2019}%
  \BibitemOpen
  \bibfield  {author} {\bibinfo {author} {\bibfnamefont {Y.}~\bibnamefont
  {Tokura}}, \bibinfo {author} {\bibfnamefont {K.}~\bibnamefont {Yasuda}}, \
  and\ \bibinfo {author} {\bibfnamefont {A.}~\bibnamefont {Tsukazaki}},\
  }\bibfield  {title} {\enquote {\bibinfo {title} {Magnetic topological
  insulators},}\ }\href {\doibase 10.1038/s42254-018-0011-5} {\bibfield
  {journal} {\bibinfo  {journal} {Nature Reviews Physics}\ }\textbf {\bibinfo
  {volume} {1}},\ \bibinfo {pages} {126--143} (\bibinfo {year}
  {2019})}\BibitemShut {NoStop}%
\bibitem [{\citenamefont {Mong}\ \emph {et~al.}(2010)\citenamefont {Mong},
  \citenamefont {Essin},\ and\ \citenamefont {Moore}}]{Mong2010}%
  \BibitemOpen
  \bibfield  {author} {\bibinfo {author} {\bibfnamefont {R.~S.~K.}\
  \bibnamefont {Mong}}, \bibinfo {author} {\bibfnamefont {A.~M.}\ \bibnamefont
  {Essin}}, \ and\ \bibinfo {author} {\bibfnamefont {J.~E.}\ \bibnamefont
  {Moore}},\ }\bibfield  {title} {\enquote {\bibinfo {title} {Antiferromagnetic
  topological insulators},}\ }\href {\doibase 10.1103/PhysRevB.81.245209}
  {\bibfield  {journal} {\bibinfo  {journal} {Phys. Rev. B}\ }\textbf {\bibinfo
  {volume} {81}},\ \bibinfo {pages} {245209} (\bibinfo {year}
  {2010})}\BibitemShut {NoStop}%
\bibitem [{\citenamefont {Wen}(2019)}]{XiaogangWen2019}%
  \BibitemOpen
  \bibfield  {author} {\bibinfo {author} {\bibfnamefont {Xiao-Gang}\
  \bibnamefont {Wen}},\ }\bibfield  {title} {\enquote {\bibinfo {title}
  {Choreographed entanglement dances: Topological states of quantum matter},}\
  }\href {\doibase 10.1126/science.aal3099} {\bibfield  {journal} {\bibinfo
  {journal} {Science}\ }\textbf {\bibinfo {volume} {363}},\ \bibinfo {pages}
  {6429} (\bibinfo {year} {2019})}\BibitemShut {NoStop}%
\bibitem [{\citenamefont {Li}\ \emph {et~al.}(2019)\citenamefont {Li},
  \citenamefont {Li}, \citenamefont {Du}, \citenamefont {Wang}, \citenamefont
  {Gu}, \citenamefont {Zhang}, \citenamefont {He}, \citenamefont {Duan},\ and\
  \citenamefont {Xu}}]{JiahengLi2019}%
  \BibitemOpen
  \bibfield  {author} {\bibinfo {author} {\bibfnamefont {J.~H.}\ \bibnamefont
  {Li}}, \bibinfo {author} {\bibfnamefont {Y.}~\bibnamefont {Li}}, \bibinfo
  {author} {\bibfnamefont {S.~Q.}\ \bibnamefont {Du}}, \bibinfo {author}
  {\bibfnamefont {Z.}~\bibnamefont {Wang}}, \bibinfo {author} {\bibfnamefont
  {B.-L.}\ \bibnamefont {Gu}}, \bibinfo {author} {\bibfnamefont {S.-C.}\
  \bibnamefont {Zhang}}, \bibinfo {author} {\bibfnamefont {K.}~\bibnamefont
  {He}}, \bibinfo {author} {\bibfnamefont {W.~H.}\ \bibnamefont {Duan}}, \ and\
  \bibinfo {author} {\bibfnamefont {Y.}~\bibnamefont {Xu}},\ }\bibfield
  {title} {\enquote {\bibinfo {title} {Intrinsic magnetic topological
  insulators in van der waals layered {MnBi$_2$Te$_4$}{-}family materials},}\
  }\href {\doibase 10.1126/sciadv.aaw5685} {\bibfield  {journal} {\bibinfo
  {journal} {Sci. Adv.}\ }\textbf {\bibinfo {volume} {5}},\ \bibinfo {pages}
  {eaaw5685} (\bibinfo {year} {2019})}\BibitemShut {NoStop}%
\bibitem [{\citenamefont {Hao}\ \emph {et~al.}()\citenamefont {Hao} \emph
  {et~al.}}]{YujieHao2019}%
  \BibitemOpen
  \bibfield  {author} {\bibinfo {author} {\bibfnamefont {Y.~J.}\ \bibnamefont
  {Hao}} \emph {et~al.},\ }\href@noop {} {\enquote {\bibinfo {title} {Gapless
  surface dirac cone in antiferromagnetomagnetic topological insulator
  {MnBi$_2$Te$_4$}},}\ }\bibinfo {note} {{a}rXiv:1907.03722}\BibitemShut
  {NoStop}%
\bibitem [{\citenamefont {Zhang}\ \emph {et~al.}(2019)\citenamefont {Zhang},
  \citenamefont {Shi}, \citenamefont {Zhu}, \citenamefont {Xing}, \citenamefont
  {Zhang},\ and\ \citenamefont {Wang}}]{DongqinZhang2019}%
  \BibitemOpen
  \bibfield  {author} {\bibinfo {author} {\bibfnamefont {D.~Q.}\ \bibnamefont
  {Zhang}}, \bibinfo {author} {\bibfnamefont {M.~J.}\ \bibnamefont {Shi}},
  \bibinfo {author} {\bibfnamefont {T.~S.}\ \bibnamefont {Zhu}}, \bibinfo
  {author} {\bibfnamefont {D.~Y.}\ \bibnamefont {Xing}}, \bibinfo {author}
  {\bibfnamefont {H.~J.}\ \bibnamefont {Zhang}}, \ and\ \bibinfo {author}
  {\bibfnamefont {J.}~\bibnamefont {Wang}},\ }\bibfield  {title} {\enquote
  {\bibinfo {title} {Topological axion states in the magnetic insulator
  {MnBi$_2$Te$_4$} with the quantized magnetoelectric effect},}\ }\href
  {\doibase 10.1103/PhysRevLett.122.206401} {\bibfield  {journal} {\bibinfo
  {journal} {Phys. Rev. Lett.}\ }\textbf {\bibinfo {volume} {122}},\ \bibinfo
  {pages} {206401} (\bibinfo {year} {2019})}\BibitemShut {NoStop}%
\bibitem [{\citenamefont {Hirahara}\ \emph {et~al.}(2017)\citenamefont
  {Hirahara} \emph {et~al.}}]{Hirahara2017}%
  \BibitemOpen
  \bibfield  {author} {\bibinfo {author} {\bibfnamefont {T.}~\bibnamefont
  {Hirahara}} \emph {et~al.},\ }\bibfield  {title} {\enquote {\bibinfo {title}
  {Large-gap magnetic topological heterostructure formed by subsurface
  incorporation of a ferromagnetic layer},}\ }\href {\doibase
  10.1021/acs.nanolett.7b00560} {\bibfield  {journal} {\bibinfo  {journal}
  {Nano Lett.}\ }\textbf {\bibinfo {volume} {17}},\ \bibinfo {pages}
  {3493--3500} (\bibinfo {year} {2017})}\BibitemShut {NoStop}%
\bibitem [{\citenamefont {Gong}\ \emph {et~al.}(2019)\citenamefont {Gong} \emph
  {et~al.}}]{YanGong2019}%
  \BibitemOpen
  \bibfield  {author} {\bibinfo {author} {\bibfnamefont {Y.}~\bibnamefont
  {Gong}} \emph {et~al.},\ }\bibfield  {title} {\enquote {\bibinfo {title}
  {Experimental realization of an intrinsic magnetic topological insulator},}\
  }\href {\doibase 10.1088/0256-307x/36/7/076801} {\bibfield  {journal}
  {\bibinfo  {journal} {Chin. Phys. Lett.}\ }\textbf {\bibinfo {volume} {36}},\
  \bibinfo {pages} {076801} (\bibinfo {year} {2019})}\BibitemShut {NoStop}%
\bibitem [{\citenamefont {Otrokov}\ \emph {et~al.}(2019)\citenamefont {Otrokov}
  \emph {et~al.}}]{Otrokov2019}%
  \BibitemOpen
  \bibfield  {author} {\bibinfo {author} {\bibfnamefont {M.~M.}\ \bibnamefont
  {Otrokov}} \emph {et~al.},\ }\bibfield  {title} {\enquote {\bibinfo {title}
  {Unique thickness-dependent properties of the van der waals interlayer
  antiferromagnet {MnBi$_2$Te$_4$} films},}\ }\href {\doibase
  10.1103/PhysRevLett.122.107202} {\bibfield  {journal} {\bibinfo  {journal}
  {Phys. Rev. Lett.}\ }\textbf {\bibinfo {volume} {122}},\ \bibinfo {pages}
  {107202} (\bibinfo {year} {2019})}\BibitemShut {NoStop}%
\bibitem [{\citenamefont {Otrokov}\ \emph {et~al.}()\citenamefont {Otrokov}
  \emph {et~al.}}]{Otrokov2018}%
  \BibitemOpen
  \bibfield  {author} {\bibinfo {author} {\bibfnamefont {M.~M.}\ \bibnamefont
  {Otrokov}} \emph {et~al.},\ }\href@noop {} {\enquote {\bibinfo {title}
  {Prediction and observation of the first antiferromagnetic topological
  insulator},}\ }\bibinfo {note} {{a}rXiv:1809.07389}\BibitemShut {NoStop}%
\bibitem [{\citenamefont {Zeugner}\ \emph {et~al.}(2019)\citenamefont {Zeugner}
  \emph {et~al.}}]{Zeugner2019}%
  \BibitemOpen
  \bibfield  {author} {\bibinfo {author} {\bibfnamefont {A.}~\bibnamefont
  {Zeugner}} \emph {et~al.},\ }\bibfield  {title} {\enquote {\bibinfo {title}
  {Chemical aspects of the candidate antiferromagnetic topological insulator
  {MnBi$_2$Te$_4$}},}\ }\href {\doibase 10.1021/acs.chemmater.8b05017}
  {\bibfield  {journal} {\bibinfo  {journal} {Chem. Mater.}\ }\textbf {\bibinfo
  {volume} {31}},\ \bibinfo {pages} {2795--2806} (\bibinfo {year}
  {2019})}\BibitemShut {NoStop}%
\bibitem [{\citenamefont {Chen}\ \emph {et~al.}()\citenamefont {Chen} \emph
  {et~al.}}]{BoChen2019}%
  \BibitemOpen
  \bibfield  {author} {\bibinfo {author} {\bibfnamefont {B.}~\bibnamefont
  {Chen}} \emph {et~al.},\ }\href@noop {} {\enquote {\bibinfo {title}
  {Searching the {Mn(Sb,Bi)$_2$Te$_4$} family of materials for the ideal
  intrinsic magnetic topological insulator},}\ }\bibinfo {note}
  {{a}rXiv:1903.09934}\BibitemShut {NoStop}%
\bibitem [{\citenamefont {Deng}\ \emph {et~al.}()\citenamefont {Deng},
  \citenamefont {Yu}, \citenamefont {Shi}, \citenamefont {Wang}, \citenamefont
  {Chen},\ and\ \citenamefont {Zhang}}]{YujunDeng2019}%
  \BibitemOpen
  \bibfield  {author} {\bibinfo {author} {\bibfnamefont {Y.~J.}\ \bibnamefont
  {Deng}}, \bibinfo {author} {\bibfnamefont {Y.~J.}\ \bibnamefont {Yu}},
  \bibinfo {author} {\bibfnamefont {M.~Z.}\ \bibnamefont {Shi}}, \bibinfo
  {author} {\bibfnamefont {J.}~\bibnamefont {Wang}}, \bibinfo {author}
  {\bibfnamefont {X.~H.}\ \bibnamefont {Chen}}, \ and\ \bibinfo {author}
  {\bibfnamefont {Y.~B.}\ \bibnamefont {Zhang}},\ }\href@noop {} {\enquote
  {\bibinfo {title} {Magnetic-field-induced quantized anomalous hall effect in
  intrinsic magnetic topological insulator {MnBi$_2$Te$_4$}},}\ }\bibinfo
  {note} {{a}rXiv:1904.11468}\BibitemShut {NoStop}%
\bibitem [{\citenamefont {Lee}\ \emph {et~al.}(2019)\citenamefont {Lee} \emph
  {et~al.}}]{SenghuatLee2018}%
  \BibitemOpen
  \bibfield  {author} {\bibinfo {author} {\bibfnamefont {S.~H.}\ \bibnamefont
  {Lee}} \emph {et~al.},\ }\bibfield  {title} {\enquote {\bibinfo {title} {Spin
  scattering and noncollinear spin structure-induced intrinsic anomalous hall
  effect in antiferromagnetic topological insulator {MnBi$_2$Te$_4$}},}\ }\href
  {\doibase 10.1103/PhysRevResearch.1.012011} {\bibfield  {journal} {\bibinfo
  {journal} {Phys. Rev. Research}\ }\textbf {\bibinfo {volume} {1}},\ \bibinfo
  {pages} {012011} (\bibinfo {year} {2019})}\BibitemShut {NoStop}%
\bibitem [{\citenamefont {Yan}\ \emph {et~al.}(2019)\citenamefont {Yan},
  \citenamefont {Zhang}, \citenamefont {Heitmann}, \citenamefont {Huang},
  \citenamefont {Chen}, \citenamefont {Cheng}, \citenamefont {Wu},
  \citenamefont {Vaknin}, \citenamefont {Sales},\ and\ \citenamefont
  {McQueeney}}]{JiaqiangYan2019}%
  \BibitemOpen
  \bibfield  {author} {\bibinfo {author} {\bibfnamefont {J.-Q.}\ \bibnamefont
  {Yan}}, \bibinfo {author} {\bibfnamefont {Q.}~\bibnamefont {Zhang}}, \bibinfo
  {author} {\bibfnamefont {T.}~\bibnamefont {Heitmann}}, \bibinfo {author}
  {\bibfnamefont {Zengle}\ \bibnamefont {Huang}}, \bibinfo {author}
  {\bibfnamefont {K.~Y.}\ \bibnamefont {Chen}}, \bibinfo {author}
  {\bibfnamefont {J.-G.}\ \bibnamefont {Cheng}}, \bibinfo {author}
  {\bibfnamefont {Weida}\ \bibnamefont {Wu}}, \bibinfo {author} {\bibfnamefont
  {D.}~\bibnamefont {Vaknin}}, \bibinfo {author} {\bibfnamefont {B.~C.}\
  \bibnamefont {Sales}}, \ and\ \bibinfo {author} {\bibfnamefont {R.~J.}\
  \bibnamefont {McQueeney}},\ }\bibfield  {title} {\enquote {\bibinfo {title}
  {Crystal growth and magnetic structure of {MnBi$_2$Te$_4$}},}\ }\href
  {\doibase 10.1103/PhysRevMaterials.3.064202} {\bibfield  {journal} {\bibinfo
  {journal} {Phys. Rev. Mater.}\ }\textbf {\bibinfo {volume} {3}},\ \bibinfo
  {pages} {064202} (\bibinfo {year} {2019})}\BibitemShut {NoStop}%
\bibitem [{\citenamefont {Aliev}\ \emph {et~al.}(2019)\citenamefont {Aliev}
  \emph {et~al.}}]{Aliev2019}%
  \BibitemOpen
  \bibfield  {author} {\bibinfo {author} {\bibfnamefont {Z.~S.}\ \bibnamefont
  {Aliev}} \emph {et~al.},\ }\bibfield  {title} {\enquote {\bibinfo {title}
  {{Novel ternary layered manganese bismuth tellurides of the
  {MnTe}-{Bi$_2$Te$_3$} system: Synthesis and crystal structure}},}\ }\href
  {\doibase 10.1016/j.jallcom.2019.03.030} {\bibfield  {journal} {\bibinfo
  {journal} {J. Alloys Compd.}\ }\textbf {\bibinfo {volume} {789}},\ \bibinfo
  {pages} {443--450} (\bibinfo {year} {2019})}\BibitemShut {NoStop}%
\bibitem [{\citenamefont {Souchay}\ \emph {et~al.}(2019)\citenamefont
  {Souchay}, \citenamefont {Nentwig}, \citenamefont {Gunther}, \citenamefont
  {Keilholz}, \citenamefont {de~Boor}, \citenamefont {Zeugner}, \citenamefont
  {Isaeva}, \citenamefont {Ruck}, \citenamefont {Wolter}, \citenamefont
  {Buchner},\ and\ \citenamefont {Oeckler}}]{Souchay2019}%
  \BibitemOpen
  \bibfield  {author} {\bibinfo {author} {\bibfnamefont {D.}~\bibnamefont
  {Souchay}}, \bibinfo {author} {\bibfnamefont {M.}~\bibnamefont {Nentwig}},
  \bibinfo {author} {\bibfnamefont {D.}~\bibnamefont {Gunther}}, \bibinfo
  {author} {\bibfnamefont {S.}~\bibnamefont {Keilholz}}, \bibinfo {author}
  {\bibfnamefont {J.}~\bibnamefont {de~Boor}}, \bibinfo {author} {\bibfnamefont
  {A.}~\bibnamefont {Zeugner}}, \bibinfo {author} {\bibfnamefont
  {A.}~\bibnamefont {Isaeva}}, \bibinfo {author} {\bibfnamefont
  {M.}~\bibnamefont {Ruck}}, \bibinfo {author} {\bibfnamefont {A.~U.~B.}\
  \bibnamefont {Wolter}}, \bibinfo {author} {\bibfnamefont {B.}~\bibnamefont
  {Buchner}}, \ and\ \bibinfo {author} {\bibfnamefont {O.}~\bibnamefont
  {Oeckler}},\ }\bibfield  {title} {\enquote {\bibinfo {title} {Layered
  manganese bismuth tellurides with {GeBi$_4$Te$_7$}- and
  {GeBi$_6$Te$_{10}$}-type structures: towards multifunctional materials},}\
  }\href {\doibase 10.1039/c9tc00979e} {\bibfield  {journal} {\bibinfo
  {journal} {J. Mater. Chem. C}\ }\textbf {\bibinfo {volume} {7}},\ \bibinfo
  {pages} {9939} (\bibinfo {year} {2019})}\BibitemShut {NoStop}%
\bibitem [{\citenamefont {Hu}\ \emph {et~al.}()\citenamefont {Hu} \emph
  {et~al.}}]{ChaoweiHu2019}%
  \BibitemOpen
  \bibfield  {author} {\bibinfo {author} {\bibfnamefont {C.-W.}\ \bibnamefont
  {Hu}} \emph {et~al.},\ }\href@noop {} {\enquote {\bibinfo {title} {A van der
  waals antiferromagnetic topological insulator with weak interlayer magnetic
  coupling},}\ }\bibinfo {note} {{a}rXiv:1905.02154}\BibitemShut {NoStop}%
\bibitem [{\citenamefont {Vidal}\ \emph {et~al.}()\citenamefont {Vidal} \emph
  {et~al.}}]{Vidal2019}%
  \BibitemOpen
  \bibfield  {author} {\bibinfo {author} {\bibfnamefont {R.~C.}\ \bibnamefont
  {Vidal}} \emph {et~al.},\ }\href@noop {} {\enquote {\bibinfo {title}
  {Topological electronic structure and intrinsic magnetization in
  {MnBi$_4$Te$_7$}: a {Bi$_2$Te$_3$}-derivative with a periodic {Mn}
  sublattice},}\ }\bibinfo {note} {{a}rXiv:1906.08394}\BibitemShut {NoStop}%
\bibitem [{\citenamefont {Cava}\ \emph {et~al.}(2013)\citenamefont {Cava},
  \citenamefont {Ji}, \citenamefont {Fuccillo}, \citenamefont {Gibson},\ and\
  \citenamefont {Hor}}]{Cava2013}%
  \BibitemOpen
  \bibfield  {author} {\bibinfo {author} {\bibfnamefont {R.~J.}\ \bibnamefont
  {Cava}}, \bibinfo {author} {\bibfnamefont {Huiwen}\ \bibnamefont {Ji}},
  \bibinfo {author} {\bibfnamefont {M.~K.}\ \bibnamefont {Fuccillo}}, \bibinfo
  {author} {\bibfnamefont {Q.~D.}\ \bibnamefont {Gibson}}, \ and\ \bibinfo
  {author} {\bibfnamefont {Y.~S.}\ \bibnamefont {Hor}},\ }\bibfield  {title}
  {\enquote {\bibinfo {title} {Crystal structure and chemistry of topological
  insulators},}\ }\href {\doibase 10.1039/C3TC30186A} {\bibfield  {journal}
  {\bibinfo  {journal} {J. Mater. Chem. C}\ }\textbf {\bibinfo {volume} {1}},\
  \bibinfo {pages} {3176--3189} (\bibinfo {year} {2013})}\BibitemShut {NoStop}%
\bibitem [{\citenamefont {Hor}\ \emph {et~al.}(2010)\citenamefont {Hor},
  \citenamefont {Roushan}, \citenamefont {Beidenkopf}, \citenamefont {Seo},
  \citenamefont {Qu}, \citenamefont {Checkelsky}, \citenamefont {Wray},
  \citenamefont {Hsieh}, \citenamefont {Xia}, \citenamefont {Xu}, \citenamefont
  {Qian}, \citenamefont {Hasan}, \citenamefont {Ong}, \citenamefont {Yazdani},\
  and\ \citenamefont {Cava}}]{Hor2010}%
  \BibitemOpen
  \bibfield  {author} {\bibinfo {author} {\bibfnamefont {Y.~S.}\ \bibnamefont
  {Hor}}, \bibinfo {author} {\bibfnamefont {P.}~\bibnamefont {Roushan}},
  \bibinfo {author} {\bibfnamefont {H.}~\bibnamefont {Beidenkopf}}, \bibinfo
  {author} {\bibfnamefont {J.}~\bibnamefont {Seo}}, \bibinfo {author}
  {\bibfnamefont {D.}~\bibnamefont {Qu}}, \bibinfo {author} {\bibfnamefont
  {J.~G.}\ \bibnamefont {Checkelsky}}, \bibinfo {author} {\bibfnamefont
  {L.~A.}\ \bibnamefont {Wray}}, \bibinfo {author} {\bibfnamefont
  {D.}~\bibnamefont {Hsieh}}, \bibinfo {author} {\bibfnamefont
  {Y.}~\bibnamefont {Xia}}, \bibinfo {author} {\bibfnamefont {S.-Y.}\
  \bibnamefont {Xu}}, \bibinfo {author} {\bibfnamefont {D.}~\bibnamefont
  {Qian}}, \bibinfo {author} {\bibfnamefont {M.~Z.}\ \bibnamefont {Hasan}},
  \bibinfo {author} {\bibfnamefont {N.~P.}\ \bibnamefont {Ong}}, \bibinfo
  {author} {\bibfnamefont {A.}~\bibnamefont {Yazdani}}, \ and\ \bibinfo
  {author} {\bibfnamefont {R.~J.}\ \bibnamefont {Cava}},\ }\bibfield  {title}
  {\enquote {\bibinfo {title} {Development of ferromagnetism in the doped
  topological insulator {Bi$_{2-x}$Mn$_x$Te$_3$}},}\ }\href {\doibase
  10.1103/PhysRevB.81.195203} {\bibfield  {journal} {\bibinfo  {journal} {Phys.
  Rev. B}\ }\textbf {\bibinfo {volume} {81}},\ \bibinfo {pages} {195203}
  (\bibinfo {year} {2010})}\BibitemShut {NoStop}%
\bibitem [{\citenamefont {Zhang}\ \emph {et~al.}(2009)\citenamefont {Zhang},
  \citenamefont {Liu}, \citenamefont {Qi}, \citenamefont {Dai}, \citenamefont
  {Fang},\ and\ \citenamefont {Zhang}}]{HaijunZhang2009}%
  \BibitemOpen
  \bibfield  {author} {\bibinfo {author} {\bibfnamefont {H.~J.}\ \bibnamefont
  {Zhang}}, \bibinfo {author} {\bibfnamefont {C.~X.}\ \bibnamefont {Liu}},
  \bibinfo {author} {\bibfnamefont {X.~L.}\ \bibnamefont {Qi}}, \bibinfo
  {author} {\bibfnamefont {X.}~\bibnamefont {Dai}}, \bibinfo {author}
  {\bibfnamefont {Z.}~\bibnamefont {Fang}}, \ and\ \bibinfo {author}
  {\bibfnamefont {S.~C.}\ \bibnamefont {Zhang}},\ }\bibfield  {title} {\enquote
  {\bibinfo {title} {Topological insulators in {Bi$_2$Se$_3$}, {Bi$_2$Te$_3$}
  and {Sb$_2$Te$_3$} with a single dirac cone on the surface},}\ }\href
  {\doibase 10.1038/nphys1270} {\bibfield  {journal} {\bibinfo  {journal}
  {Nature Phys.}\ }\textbf {\bibinfo {volume} {5}},\ \bibinfo {pages}
  {438-442} (\bibinfo {year} {2009})}\BibitemShut {NoStop}%
\bibitem [{\citenamefont {Chen}\ \emph {et~al.}(2009)\citenamefont {Chen} \emph
  {et~al.}}]{YulinChen2009}%
  \BibitemOpen
  \bibfield  {author} {\bibinfo {author} {\bibfnamefont {Y.~L.}\ \bibnamefont
  {Chen}} \emph {et~al.},\ }\bibfield  {title} {\enquote {\bibinfo {title}
  {Experimental realization of a three-dimensional topological insulator,
  {Bi$_2$Te$_3$}},}\ }\href {\doibase 10.1126/science.1173034} {\bibfield
  {journal} {\bibinfo  {journal} {Science}\ }\textbf {\bibinfo {volume}
  {325}},\ \bibinfo {pages} {178--182} (\bibinfo {year} {2009})}\BibitemShut
  {NoStop}%
\bibitem [{\citenamefont {Atuchin}\ \emph {et~al.}(2012)\citenamefont
  {Atuchin}, \citenamefont {Gavrilova}, \citenamefont {Kokh}, \citenamefont
  {Kuratieva}, \citenamefont {Pervukhina},\ and\ \citenamefont
  {Surovtsev}}]{Atuchin2012}%
  \BibitemOpen
  \bibfield  {author} {\bibinfo {author} {\bibfnamefont {V.V.}\ \bibnamefont
  {Atuchin}}, \bibinfo {author} {\bibfnamefont {T.A.}\ \bibnamefont
  {Gavrilova}}, \bibinfo {author} {\bibfnamefont {K.A.}\ \bibnamefont {Kokh}},
  \bibinfo {author} {\bibfnamefont {N.V.}\ \bibnamefont {Kuratieva}}, \bibinfo
  {author} {\bibfnamefont {N.V.}\ \bibnamefont {Pervukhina}}, \ and\ \bibinfo
  {author} {\bibfnamefont {N.V.}\ \bibnamefont {Surovtsev}},\ }\bibfield
  {title} {\enquote {\bibinfo {title} {Structural and vibrational properties of
  pvt grown {Bi$_2$Te$_3$} microcrystals},}\ }\href {\doibase
  10.1016/j.ssc.2012.04.007} {\bibfield  {journal} {\bibinfo  {journal} {Solid
  State Commun.}\ }\textbf {\bibinfo {volume} {152}},\ \bibinfo {pages}
  {1119--1122} (\bibinfo {year} {2012})}\BibitemShut {NoStop}%
\bibitem [{\citenamefont {Ye}\ \emph {et~al.}(2018)\citenamefont {Ye},
  \citenamefont {Liu}, \citenamefont {Whitfield}, \citenamefont {Osborn},\ and\
  \citenamefont {Rosenkranz}}]{FengYe2018}%
  \BibitemOpen
  \bibfield  {author} {\bibinfo {author} {\bibfnamefont {F.}~\bibnamefont
  {Ye}}, \bibinfo {author} {\bibfnamefont {Y.}~\bibnamefont {Liu}}, \bibinfo
  {author} {\bibfnamefont {R.}~\bibnamefont {Whitfield}}, \bibinfo {author}
  {\bibfnamefont {R.}~\bibnamefont {Osborn}}, \ and\ \bibinfo {author}
  {\bibfnamefont {S.}~\bibnamefont {Rosenkranz}},\ }\bibfield  {title}
  {\enquote {\bibinfo {title} {Implementation of cross correlation for energy
  discrimination on the time-of-flight spectrometer {CORELLI}},}\ }\href
  {\doibase 10.1107/S160057671800403X} {\bibfield  {journal} {\bibinfo
  {journal} {J. Appl. Crystallogr.}\ }\textbf {\bibinfo {volume} {51}},\
  \bibinfo {pages} {315--322} (\bibinfo {year} {2018})}\BibitemShut {NoStop}%
\bibitem [{sup()}]{suppl}%
  \BibitemOpen
  \href@noop {} {}\bibinfo {note} {See Supplemental Material for experimental
  methods and the description of the crystal structure refinement and magnetic
  structure determination.}\BibitemShut {Stop}%
\bibitem [{\citenamefont {Chakoumakos}\ \emph {et~al.}(2011)\citenamefont
  {Chakoumakos}, \citenamefont {Cao}, \citenamefont {Ye}, \citenamefont
  {Stoica}, \citenamefont {Popovici}, \citenamefont {Sundaram}, \citenamefont
  {Zhou}, \citenamefont {Hicks}, \citenamefont {Lynn},\ and\ \citenamefont
  {Riedel}}]{HB3A}%
  \BibitemOpen
  \bibfield  {author} {\bibinfo {author} {\bibfnamefont {B.~C.}\ \bibnamefont
  {Chakoumakos}}, \bibinfo {author} {\bibfnamefont {H.~B.}\ \bibnamefont
  {Cao}}, \bibinfo {author} {\bibfnamefont {F.}~\bibnamefont {Ye}}, \bibinfo
  {author} {\bibfnamefont {A.~D.}\ \bibnamefont {Stoica}}, \bibinfo {author}
  {\bibfnamefont {M.}~\bibnamefont {Popovici}}, \bibinfo {author}
  {\bibfnamefont {M.}~\bibnamefont {Sundaram}}, \bibinfo {author}
  {\bibfnamefont {W.}~\bibnamefont {Zhou}}, \bibinfo {author} {\bibfnamefont
  {J.~S.}\ \bibnamefont {Hicks}}, \bibinfo {author} {\bibfnamefont {G.~W.}\
  \bibnamefont {Lynn}}, \ and\ \bibinfo {author} {\bibfnamefont {R.~A.}\
  \bibnamefont {Riedel}},\ }\bibfield  {title} {\enquote {\bibinfo {title}
  {Symmetry-based computational tools for magnetic crystallography},}\ }\href
  {\doibase 10.1107/S0021889811012301} {\bibfield  {journal} {\bibinfo
  {journal} {J. Appl. Crystallogr.}\ }\textbf {\bibinfo {volume} {44}},\
  \bibinfo {pages} {655} (\bibinfo {year} {2011})}\BibitemShut {NoStop}%
\bibitem [{\citenamefont {Cao}\ \emph {et~al.}(2019)\citenamefont {Cao},
  \citenamefont {Chakoumakos}, \citenamefont {Andrews}, \citenamefont {Wu},
  \citenamefont {Riedel}, \citenamefont {Hodges}, \citenamefont {Zhou},
  \citenamefont {Gregory}, \citenamefont {Haberi}, \citenamefont {Molaison},\
  and\ \citenamefont {Lynn}}]{DEMAND}%
  \BibitemOpen
  \bibfield  {author} {\bibinfo {author} {\bibfnamefont {H.~B.}\ \bibnamefont
  {Cao}}, \bibinfo {author} {\bibfnamefont {B.~C.}\ \bibnamefont
  {Chakoumakos}}, \bibinfo {author} {\bibfnamefont {K.~M.}\ \bibnamefont
  {Andrews}}, \bibinfo {author} {\bibfnamefont {Y.}~\bibnamefont {Wu}},
  \bibinfo {author} {\bibfnamefont {R.~A.}\ \bibnamefont {Riedel}}, \bibinfo
  {author} {\bibfnamefont {J.}~\bibnamefont {Hodges}}, \bibinfo {author}
  {\bibfnamefont {W.~D.}\ \bibnamefont {Zhou}}, \bibinfo {author}
  {\bibfnamefont {R.}~\bibnamefont {Gregory}}, \bibinfo {author} {\bibfnamefont
  {B.}~\bibnamefont {Haberi}}, \bibinfo {author} {\bibfnamefont
  {J.}~\bibnamefont {Molaison}}, \ and\ \bibinfo {author} {\bibfnamefont
  {G.~W.}\ \bibnamefont {Lynn}},\ }\bibfield  {title} {\enquote {\bibinfo
  {title} {Demand, a dimensional extreme magnetic neutron diffractometer at the
  high flux isotope reactor},}\ }\href {\doibase 10.3390/cryst9010005}
  {\bibfield  {journal} {\bibinfo  {journal} {Crystals}\ }\textbf {\bibinfo
  {volume} {9}},\ \bibinfo {pages} {5} (\bibinfo {year} {2019})}\BibitemShut
  {NoStop}%
\bibitem [{\citenamefont {Rodr\'{\i}guez-Carvajal}(1993)}]{Fullprof1993}%
  \BibitemOpen
  \bibfield  {author} {\bibinfo {author} {\bibfnamefont {J.}~\bibnamefont
  {Rodr\'{\i}guez-Carvajal}},\ }\bibfield  {title} {\enquote {\bibinfo {title}
  {Recent advances in magnetic structure determination by neutron powder
  diffraction},}\ }\href {\doibase 10.1016/0921-4526(93)90108-I} {\bibfield
  {journal} {\bibinfo  {journal} {Phys. B Condens. Matter}\ }\textbf {\bibinfo
  {volume} {192}},\ \bibinfo {pages} {55} (\bibinfo {year} {1993})}\BibitemShut
  {NoStop}%
\bibitem [{\citenamefont {Lee}\ \emph {et~al.}(2013)\citenamefont {Lee},
  \citenamefont {Kim}, \citenamefont {Park}, \citenamefont {Chung},
  \citenamefont {Lim}, \citenamefont {Seo},\ and\ \citenamefont
  {Park}}]{DongsunLee2013}%
  \BibitemOpen
  \bibfield  {author} {\bibinfo {author} {\bibfnamefont {D.~S.}\ \bibnamefont
  {Lee}}, \bibinfo {author} {\bibfnamefont {T.-H.}\ \bibnamefont {Kim}},
  \bibinfo {author} {\bibfnamefont {C.-H.}\ \bibnamefont {Park}}, \bibinfo
  {author} {\bibfnamefont {C.-Y.}\ \bibnamefont {Chung}}, \bibinfo {author}
  {\bibfnamefont {Y.~S.}\ \bibnamefont {Lim}}, \bibinfo {author} {\bibfnamefont
  {W.-S.}\ \bibnamefont {Seo}}, \ and\ \bibinfo {author} {\bibfnamefont
  {H.-H.}\ \bibnamefont {Park}},\ }\bibfield  {title} {\enquote {\bibinfo
  {title} {Crystal structure{,} properties and nanostructuring of a new layered
  chalcogenide semiconductor{,} {Bi$_2$MnTe$_4$}},}\ }\href {\doibase
  10.1039/C3CE40643A} {\bibfield  {journal} {\bibinfo  {journal}
  {CrystEngComm}\ }\textbf {\bibinfo {volume} {15}},\ \bibinfo {pages}
  {5532--5538} (\bibinfo {year} {2013})}\BibitemShut {NoStop}%
\bibitem [{\citenamefont {Campbell}\ \emph {et~al.}(2006)\citenamefont
  {Campbell}, \citenamefont {Stokes}, \citenamefont {E.},\ and\ \citenamefont
  {Hatch}}]{ISODISTORT}%
  \BibitemOpen
  \bibfield  {author} {\bibinfo {author} {\bibfnamefont {B.~J.}\ \bibnamefont
  {Campbell}}, \bibinfo {author} {\bibfnamefont {H.~T.}\ \bibnamefont
  {Stokes}}, \bibinfo {author} {\bibfnamefont {Tanner~D.}\ \bibnamefont {E.}},
  \ and\ \bibinfo {author} {\bibfnamefont {D.~M.}\ \bibnamefont {Hatch}},\
  }\bibfield  {title} {\enquote {\bibinfo {title} {Isodisplace: An internet
  tool for exploring structural distortions},}\ }\href {\doibase
  10.1107/S0021889806014075} {\bibfield  {journal} {\bibinfo  {journal} {J.
  Appl. Crystallogr.}\ }\textbf {\bibinfo {volume} {39}},\ \bibinfo {pages}
  {607} (\bibinfo {year} {2006})}\BibitemShut {NoStop}%
\bibitem [{\citenamefont {Perez-Mato}\ \emph {et~al.}(2015)\citenamefont
  {Perez-Mato}, \citenamefont {Gallego}, \citenamefont {Tasci}, \citenamefont
  {Elcoro}, \citenamefont {de~la Flor},\ and\ \citenamefont {Aroyo}}]{bilbao}%
  \BibitemOpen
  \bibfield  {author} {\bibinfo {author} {\bibfnamefont {J.~M.}\ \bibnamefont
  {Perez-Mato}}, \bibinfo {author} {\bibfnamefont {S.~V.}\ \bibnamefont
  {Gallego}}, \bibinfo {author} {\bibfnamefont {E.~S.}\ \bibnamefont {Tasci}},
  \bibinfo {author} {\bibfnamefont {L.}~\bibnamefont {Elcoro}}, \bibinfo
  {author} {\bibfnamefont {G.}~\bibnamefont {de~la Flor}}, \ and\ \bibinfo
  {author} {\bibfnamefont {M.~I.}\ \bibnamefont {Aroyo}},\ }\bibfield  {title}
  {\enquote {\bibinfo {title} {Symmetry-based computational tools for magnetic
  crystallography},}\ }\href {\doibase 10.1146/annurev-matsci-070214-021008}
  {\bibfield  {journal} {\bibinfo  {journal} {Ann. Rev. Mater. Res.}\ }\textbf
  {\bibinfo {volume} {45}},\ \bibinfo {pages} {217} (\bibinfo {year}
  {2015})}\BibitemShut {NoStop}%
\bibitem [{\citenamefont {Brush}(1967)}]{Brush1967}%
  \BibitemOpen
  \bibfield  {author} {\bibinfo {author} {\bibfnamefont {S.~G.}\ \bibnamefont
  {Brush}},\ }\bibfield  {title} {\enquote {\bibinfo {title} {History of the
  lenz-ising model},}\ }\href {\doibase 10.1103/RevModPhys.39.883} {\bibfield
  {journal} {\bibinfo  {journal} {Rev. Mod. Phys.}\ }\textbf {\bibinfo {volume}
  {39}},\ \bibinfo {pages} {883--893} (\bibinfo {year} {1967})}\BibitemShut
  {NoStop}%
\bibitem [{\citenamefont {Birgeneau}\ \emph {et~al.}(1970)\citenamefont
  {Birgeneau}, \citenamefont {Guggenheim},\ and\ \citenamefont
  {Shirane}}]{Birgeneau1970}%
  \BibitemOpen
  \bibfield  {author} {\bibinfo {author} {\bibfnamefont {R.~J.}\ \bibnamefont
  {Birgeneau}}, \bibinfo {author} {\bibfnamefont {H.~J.}\ \bibnamefont
  {Guggenheim}}, \ and\ \bibinfo {author} {\bibfnamefont {G.}~\bibnamefont
  {Shirane}},\ }\bibfield  {title} {\enquote {\bibinfo {title} {Neutron
  scattering investigation of phase transitions and magnetic correlations in
  the two-dimensional antiferromagnets {K$_2$NiF$_4$}, {Rb$_2$MnF$_4$},
  {Rb$_2$FeF$_4$}},}\ }\href {\doibase 10.1103/PhysRevB.1.2211} {\bibfield
  {journal} {\bibinfo  {journal} {Phys. Rev. B}\ }\textbf {\bibinfo {volume}
  {1}},\ \bibinfo {pages} {2211--2230} (\bibinfo {year} {1970})}\BibitemShut
  {NoStop}%
\bibitem [{\citenamefont {Wildes}\ \emph {et~al.}(2006)\citenamefont {Wildes},
  \citenamefont {R\o{}nnow}, \citenamefont {Roessli}, \citenamefont {Harris},\
  and\ \citenamefont {Godfrey}}]{Wildes2006}%
  \BibitemOpen
  \bibfield  {author} {\bibinfo {author} {\bibfnamefont {A.~R.}\ \bibnamefont
  {Wildes}}, \bibinfo {author} {\bibfnamefont {H.~M.}\ \bibnamefont
  {R\o{}nnow}}, \bibinfo {author} {\bibfnamefont {B.}~\bibnamefont {Roessli}},
  \bibinfo {author} {\bibfnamefont {M.~J.}\ \bibnamefont {Harris}}, \ and\
  \bibinfo {author} {\bibfnamefont {K.~W.}\ \bibnamefont {Godfrey}},\
  }\bibfield  {title} {\enquote {\bibinfo {title} {Static and dynamic critical
  properties of the quasi-two-dimensional antiferromagnet {MnPS$_3$}},}\ }\href
  {\doibase 10.1103/PhysRevB.74.094422} {\bibfield  {journal} {\bibinfo
  {journal} {Phys. Rev. B}\ }\textbf {\bibinfo {volume} {74}},\ \bibinfo
  {pages} {094422} (\bibinfo {year} {2006})}\BibitemShut {NoStop}%
\bibitem [{\citenamefont {Bos}\ \emph {et~al.}(2007)\citenamefont {Bos},
  \citenamefont {Zandbergen}, \citenamefont {Lee}, \citenamefont {Ong},\ and\
  \citenamefont {Cava}}]{Bos2007}%
  \BibitemOpen
  \bibfield  {author} {\bibinfo {author} {\bibfnamefont {J.~W.~G.}\
  \bibnamefont {Bos}}, \bibinfo {author} {\bibfnamefont {H.~W.}\ \bibnamefont
  {Zandbergen}}, \bibinfo {author} {\bibfnamefont {M.-H.}\ \bibnamefont {Lee}},
  \bibinfo {author} {\bibfnamefont {N.~P.}\ \bibnamefont {Ong}}, \ and\
  \bibinfo {author} {\bibfnamefont {R.~J.}\ \bibnamefont {Cava}},\ }\bibfield
  {title} {\enquote {\bibinfo {title} {Structures and thermoelectric properties
  of the infinitely adaptive series {(Bi$_2$)$_m$(Bi$_2$Te$_3$)$_n$}},}\ }\href
  {\doibase 10.1103/PhysRevB.75.195203} {\bibfield  {journal} {\bibinfo
  {journal} {Phys. Rev. B}\ }\textbf {\bibinfo {volume} {75}},\ \bibinfo
  {pages} {195203} (\bibinfo {year} {2007})}\BibitemShut {NoStop}%
\bibitem [{DOE()}]{DOE}%
  \BibitemOpen
  \href@noop {} {}\bibinfo {note}
  {Https://www.energy.gov/downloads/doe-public-access-plan}\BibitemShut
  {NoStop}%
\end{thebibliography}
\end{document}